\documentclass[12pt]{article}
\usepackage{amsmath,amsfonts,amssymb}
\usepackage[nosort]{cite}
\usepackage{hyperref}
\usepackage{graphicx}
\usepackage{color}
\usepackage{epsfig}
\usepackage{psfrag}
\usepackage{tikz}
\usepackage{slashed}
\usepackage{youngtab}
\usepackage{ulem}
\usepackage{comment}

\relax

\makeatletter\@addtoreset{equation}{section}\makeatother

\DeclareMathOperator{\Tr}{Tr}

\DeclareMathOperator{\arctanh}{arctanh}

\DeclareMathOperator{\cn}{cn}

\DeclareMathOperator{\ch}{ch}

\def\bZ {\mathbb{Z}}

\newcommand{\beq}{\begin{equation}}
\newcommand{\eeq}{\end{equation}}
\newcommand{\bal}{\begin{equation}\begin{aligned}}
\newcommand{\eal}{\end{aligned}\end{equation}}
\newcommand{\half}{\frac{1}{2}}

\newcommand{\eqn}[1]{(\ref{#1})}

\newcommand{\address}[1]{\vbox{\center\em#1}}
\renewcommand{\title}[1]{\vbox{\center\huge{#1}}\vspace{5mm}}

\newcommand{\cN}{{\mathcal N}}

\newcommand{\cO}{{\mathcal O}}

\newcommand{\cI}{{\mathcal I}}

\begin{document}
\bibliographystyle{utphys2}

\begin{titlepage}
\begin{center}
\phantom{ }

\vspace{5mm}

\title{The $\cN=4$ Schur index with Polyakov loops}
\vspace{3mm}

\renewcommand{\thefootnote}{$\alph{footnote}$}

Nadav Drukker%
\footnote{\href{mailto:nadav.drukker@gmail.com}{\tt nadav.drukker@gmail.com}} 

\address{
Department of Mathematics, King's College London \\
The Strand, WC2R 2LS, London, UK}

\renewcommand{\thefootnote}{\arabic{footnote}}
\setcounter{footnote}{0}

\end{center}

\vskip5mm

\abstract{
\normalsize
\noindent
Recently the Schur index of $\cN=4$ SYM was evaluated in closed form to all orders 
including exponential corrections in the large $N$ expansion and for fixed finite $N$. 
This was achieved by identifying the matrix model which calculates the index with 
the partition function of a system of free fermions on a circle. The index can be 
enriched by the inclusion of loop operators and the case of Wilson loops is particularly 
easy, as it amounts to inserting extra characters into the matrix model. The Fermi-gas 
approach is applied here to this problem, the formalism is explored and explicit results 
at large $N$ are found for the fundamental as well as a few other symmetric and 
antisymmetric representations.
}

\end{titlepage}

\setcounter{tocdepth}{2} 
\addtolength{\parskip}{-.4mm}
\addtolength{\parskip}{.4mm}

\section{Introduction}
\label{sec:intro}

The superconformal index of four dimensional field theories 
\cite{Romelsberger2006,Kinney2007} provides an interesting 
probe to the spectrum of such theories. It can be defined by a weighted sum with 
alternating signs for bosons and fermions and extra fugacities for global symmetries. 
This sum over all states can be combined from the ``single letter index'' contribution 
of a single field to that of gauge invariant composite operators, organized by specific 
factors for the different multiplets. In particular, the Schur index which counts states 
preserving double the minimal amount of supersymmetery is an ``unrefinement'' of the 
usual index. It is achieved by eliminating 
some of the global charge fugacities and can be expressed as an elliptic matrix model --- 
where the measure and interaction between eigenvalues are all elliptic functions.

Recently it was realized that this matrix model in the case of $\cN=4$ SYM 
can be mapped to the problem of one dimensional fermions on a circle with 
no interaction nor any potential \cite{BDF1}, as reviewed briefly below 
(the generalization to circular quiver theories is done in \cite{BDF2}). This problem 
can be solved in closed form giving the exact large $N$ expansion of the index 
including all exponentially suppressed corrections as well as finite $N$ 
expressions in terms of derivatives of Jacobi theta functions or 
complete elliptic integrals.

The index can be viewed as the partition function of the theory on $S^3\times S^1$ 
with supersymmetry preserving boundary conditions around the circle. As such, one can 
consider the insertion of loop operators wrapping the non-contractible circle, also 
known as Polyakov loops. These can be electric lines ({\em i.e.}, Wilson loops), 
magnetic ('t Hoof loops) or dyonic. The effect of such insertions on the matrix model 
of the index was studied in \cite{Gang:2012yr} 
(see also \cite{takuya,diego}). One important point is that due to the Gauss law 
on the compact $S^3$, the total charge carried by the line operators must vanish. 
In particular, the simplest non-trivial insertion is a pair of lines carrying opposite 
charges one at each of the north and south poles of $S^3$. The purpose of this note 
is to study the case of Wilson loops insertions using the newly discovered 
formalism of \cite{BDF1}.

\subsection{The Schur index as a Fermi gas}

The Schur index of $\cN=4$ SYM on $S^3\times S^1$ with gauge group $U(N)$ is 
given by the matrix model 
\cite{Romelsberger2006,Kinney2007,Dolan2009,Gadde2010,macdonald,Razamat2012,BDF1}
\beq
\label{4dvec}
\cI(N)
=\frac{q^{-N^2/4}\eta^{3N}(\tau)}{N!\pi^N}
\int_0^\pi d^N \alpha\,
\frac{\prod_{i <j} \vartheta_1^2( \alpha_i-\alpha_j)}
{\prod_{i,j} \vartheta_4(\alpha_i-\alpha_j)}\,.
\eeq
Here $q=e^{i\pi\tau}$ is the one fugacity which remains in the restriction to the Schur 
index.\footnote{%
Following the conventions of \cite{BDF1}, which are slightly different than the 
rest of the literature.}
$\eta$ is the Dedekind function and $\vartheta_i(z)$ are 
Jacobi theta functions with the nome $q$ suppressed (see Appendix~\ref{app:theta}). 
When the argument $z$ is also omitted (as in \eqn{SYMfermi} below), 
this is $\vartheta_i(0)$.

Using an elliptic determinant identity 
\cite{Frobenius1879,Frobenius1882,Krattenthaler2005,BDF1} this can be written as
\beq
\label{SYMfermi}
\cI(N) =\frac{q^{-N^2/4 }}{\Delta_N}\,Z(N)\,,
\qquad
Z(N)= \frac{1}{N!} \sum_{\sigma \in S_N} 
(-1)^\sigma \int_0^\pi
d^N\alpha \prod_{i=1}^N 
\frac{\vartheta_2^2}{2\pi}\cn \big( (\alpha_i - \alpha_{\sigma(i)} )\vartheta_3^2 \big)\,.
\eeq
Here $\cn(z)\equiv\cn(z,k^2)$ is a Jacobi elliptic function with the usual modulus associated 
to $q$, given by $k=\vartheta_2^2/\vartheta_3^2$ and the normalisation is
\beq
\label{delta}
\Delta_N 
= \begin{cases} 1\,, &\text{$N$ even} ,
\\
{\vartheta_2/\vartheta_3} \,, & \text{$N$ odd} \,.
\end{cases}
\eeq

Equation \eqn{SYMfermi} has the form of the partition function of $N$ 
\emph{free} fermions on a circle. 
The Fermi gas partition function is completely determined by the spectral traces
\bal
\label{Zldef}
Z_\ell&= \Tr (\rho_0^\ell) 
=\frac{1}{\pi^N} \int_0^\pi d \alpha_1 \dots d\alpha_\ell \, 
\rho_0\left( \alpha_1, \alpha_2 \right) \dots \rho_0 \left( \alpha_\ell, \alpha_1 \right) \, ,
\\
\rho_0\big(\alpha,\alpha'\big)
&=\frac{\vartheta_2^2}{2}\cn \big((\alpha - \alpha')\vartheta_3^2 \big)
=\sum_{p\in2\bZ+1}\frac{e^{ip( \alpha - \alpha') }}{q^{p/2}+q^{-p/2}}\,,
\eal
where the last identity uses the Fourier expansion of the $\cn$ function.
It is convenient to define
\beq
\ch p\equiv 2\cosh\frac{i\pi\tau p}{2}=q^{p/2}+q^{-p/2}\,.
\eeq
It is then easy to perform the integrals in \eqn{Zldef} to find
\beq
\label{Zlexp}
Z_\ell
=\sum_{p\in2\bZ+1}\frac{1}{\ch^\ell p}\,.
\eeq

The approach employed in \cite{BDF1} (following \cite{MF2} who studied ABJM theory) 
was to introduce a fugacity $\kappa$ and consider the 
sum over the partition function of the fermions associated to theories with arbitrary rank $N$.
This gives the grand canonical partition function
\bal
\label{grandpartition}
\Xi(\kappa) = 1 + \sum_{N=1}^\infty Z(N) \kappa^N\,.
\eal
The result is a Fredholm determinant of a very simple form
\beq
\label{SUNXi}
\Xi(\kappa) = \exp\bigg({-}\sum_{\ell= 1}^\infty \frac{(-\kappa)^\ell Z_\ell}{\ell}\bigg)
=\det(1+\kappa\rho_0)= \prod_{p \in 2\bZ+1} \left( 1 + 
\frac{\kappa}{\ch p} \right).
\eeq
This product turns out to be expressible in terms of theta functions 
\cite{BDF1}
\beq
\label{xi-exact}
\Xi(\kappa) 
= \prod_{p=1}^\infty \left( \frac{1 + q^{2p-1} + \kappa q^{-(p-1/2) } }
{1 + q^{2p-1} }\right)^2 
= \frac{1}{\vartheta_4}
\left(\vartheta_3\Big(\arccos\frac{\kappa}{2}\Big)
+\frac{\vartheta_2}{\vartheta_3}\vartheta_2\Big(\arccos\frac{\kappa}{2}\Big)\right).
\eeq

\subsection{The Schur index with Polyakov loops}

As stated above, the purpose of this note is to study the index in the presence 
of line operators at the north and south poles. 
For Wilson loops in conjugate representations $R$ and $\bar R$ the matrix integral 
in \eqn{4dvec} gets modified by the insertion of the characters of those representations
\beq
\label{WL-def}
\cI_R(N)
=\frac{1}{\cI(N)}\frac{q^{-N^2/4}\eta^{3N}(\tau)}{N!\pi^N}
\int_0^\pi d^N \alpha\,
\Tr_R(e^{2i\alpha}) \Tr_{\bar R}(e^{2i\alpha})
\frac{\prod_{i <j} \vartheta_1^2( \alpha_i-\alpha_j)}
{\prod_{i,j} \vartheta_4(\alpha_i-\alpha_j)}\,.
\eeq
Note that this expression is normalized with the index in the absence of the Wilson loop, 
which simplifies some of the steps to come. In particular the prefactor relating 
$\cI(N)$ and $Z(N)$ 
in \eqn{SYMfermi} does not affect this quantity, so after applying the determinant 
identity as before, the matrix model becomes ({\it c.f.}, \eqn{SYMfermi})
\beq
\label{WL}
\cI_R(N)
=\frac{1}{Z(N)N!} \sum_{\sigma \in S_N} 
(-1)^\sigma \int_0^\pi
\frac{d^N\alpha}{\pi^N} \,\Tr_R(e^{2i\alpha}) \Tr_{\bar R}(e^{2i\alpha})\,\prod_{i=1}^N 
\rho_0 (\alpha_i,\alpha_{\sigma(i)} )\,.
\eeq

For the fundamental representation the Wilson loop (in the canonical ensemble) is
\beq
\label{fund}
\cI_\Box(N)
=\frac{1}{Z(N)N!} \sum_{\sigma \in S_N} 
(-1)^\sigma \int_0^\pi
\frac{d^N\alpha}{\pi^N} \,\sum_{j,k=1}^Ne^{2i\alpha_j}e^{-2i\alpha_k}\,\prod_{i=1}^N 
\rho_0 (\alpha_i,\alpha_{\sigma(i)} )\,.
\eeq
Or course as far as the matrix model is concerned, this insertion is actually the product 
of the fundamental and antifundamental, which is the direct sum of the identity 
and adjoint. Each of the latter have a non vanishing VEV in this matrix model. 
Still, in keeping with the picture of the index with two Wilson loop insertions, 
this is labeled here as the fundamental representation. Likewise when 
discussing antisymmetric representations below, the matrix model insertion is the 
product of two antisymmetrics. For the $n$\textsuperscript{th} antisymmetric 
representation, the product is a direct sum of  $n+1$ representations with $N$ 
boxes and two columns where the second column's height is less or equal 
to $n$ (including zero, the identity representation).

For the symmetric representation the product is reducible to the sum of 
$n+1$ representations with $i$ filled columns, $n-i$ columns of height $N-1$ and 
the same number of columns of unit height. Again, those are denoted below by the 
symmetric Young diagram, and not by the product representation.

The Fermi-gas approach to solving the matrix models describing supersymmetric 
field theories in three dimensions was pioneered in \cite{MF2}. This was 
generalized to allow for Wilson loop operators in
\cite{Klemm:2012ii,Hatsuda:2013yua,Hirano:2014bia} 
(see also \cite{Ouyang:2015hta}). 
A lot of the techniques used below are taken from these papers. 
As in some of those papers, it proves useful to define the Wilson loop also 
in the grand canonical ensemble by
\beq
\label{grand-R}
W_R(\kappa)
=\frac{1}{\Xi(\kappa)}\sum_{N=1}^\infty \cI_R(N)Z(N)\kappa^N\,.
\eeq

It should be noted that in addition to Wilson loops, four dimensional field theories have 
BPS 't~Hooft loops (and dyonic ones). Their effect on the matrix model was also studied 
in \cite{Gang:2012yr}. In $\cN=4$ SYM 't~Hooft loops should be $S$-dual to Wilson loops, 
and therefore the index with 't~Hooft loops and with Wilson loops should be equal. This 
has not been demonstrated in \cite{Gang:2012yr} nor is it pursued here. The results presented 
below apply to 't~Hooft loops {\it assuming} $S$-duality and it would be desirable to have 
a direct calculation to demonstrate this equivalence.

In the next section some formalism is developed to write down the index with Wilson loop 
insertions in the grand canonical ensemble. The resulting expressions for the first few 
antisymmetric and symmetric representations are given as an infinite sum over an 
explicit $\kappa$ dependent function. The following section studies those sums in the 
large $\kappa$ limit, where they can be approximated by an integral. The leading large 
$N$ result can then be derived for the first few antisymmetric and symmetric representations. 
A simple pattern emerges in this calculation and it is conjectured to apply to higher dimensional 
representations too. 
Finally, in Section~\ref{sec:finiteN} the sums arising at finite $N$ are explored and 
the case of the fundamental representation for $N=2$ is evaluated in closed form.

\section{Generating functions}
\label{sec:gen}

\subsection{Antisymmetric representations}

To study the Wilson loops in the grand canonical ensemble and apply the Fermi-gas 
formalism it is easier to include a determinant, rather than a trace, as done for 
ABJM theory in \cite{Hatsuda:2013yua}. 
This is easy to implement, since the characters of the antisymmetric 
representation are the symmetric 
polynomials generated by $\prod_{j=1}^N\left(1+se^{2i\alpha_j}\right)$. Including a second 
conjugate representation one finds
\bal
\label{anti-gen}
\prod_{j=1}^N\left(1+se^{2i\alpha_j}\right)\left(1+te^{-2i\alpha_j}\right)
&=1+\sum_{j=1}^N\left(se^{2i\alpha_j}+te^{-2i\alpha_j}\right)
+st\sum_{j,k=1}^Ne^{2i(\alpha_j-\alpha_k)}
\\&\quad{}
+\sum_{j<k}\left(s^2e^{2i(\alpha_j+\alpha_k)}+t^2e^{-2i(\alpha_j+\alpha_k)}\right)
+\cdots\,.
\eal
At order $st$ this is indeed the product of the traces in the fundamental 
and in the antifundamental representations \eqn{fund}.

As mentioned before, the index vanishes unless two conjugate representations 
are inserted, so only the terms of equal powers of $s$ and $t$ survive. This is also 
evident in the calculation below. The inclusion of the generating function 
of the Wilson loops in the partition function amounts to replacing the density operator 
$\rho_0$ \eqn{Zldef} with
\beq
\rho_A\equiv(1+se^{2i\alpha})(1+te^{-2i\alpha})\rho_0\,.
\eeq

The generating function of the Wilson loops in the antisymmetric representations in the 
grand canonical ensemble is then
\beq
\label{grand-anti}
1+\sum_{k=1}^\infty(st)^kW_{\text{asym}^k}(\kappa)
=\frac{\det\left(1+\kappa\rho_A\right)}{\det\left(1+\kappa\rho_0\right)}
=\det\left(1+X_AR\right).
\eeq
where
\beq
\label{XAR}
X_A=se^{2i\alpha}+te^{-2i\alpha}+st\,,
\qquad
R=\frac{\kappa\rho_0}{1+\kappa\rho_0}=\frac{\kappa}{\kappa+\ch p}\,.
\eeq
It is useful to write the determinant as
\bal
\det(1+X_AR)
=\exp\left[\sum_{n=1}^\infty\frac{(-1)^{n+1}}{n}\Tr(X_AR)^n\right].
\eal
In these expressions one should think of $\alpha$ and $p$ as conjugate position-momentum 
operators (periodic and discrete, repectively). 
If one commutes the $\alpha$ dependent terms in $X_A$ through $R$, 
then in the momentum basis 
\beq
\label{Rn}
R_{n}\equiv e^{-2in\alpha}Re^{2in\alpha}=\frac{\kappa}{\kappa+\ch(p+2n)}\,,
\eeq
and for simplicity denote $R_\pm\equiv R_{\pm1}$. This allows to write explicit normal ordered
expressions for $(X_AR)^n$. For example
\bal
\label{someXA^n}
(X_AR)^2&=s^2e^{4i\alpha}RR_+
+s^2te^{2i\alpha}R(R+R_+)
+stR(R_-+R_+)
+s^2t^2 R^2
\\&\quad{}
+st^2e^{-2i\alpha}R(R+R_-)
+t^2e^{-4i\alpha}RR_-\,.
\eal
Clearly the terms with $s^lt^m$ come with $e^{2i(l-m)\alpha}$, whose trace is 
zero unless $l=m$ (also when multiplying a function of $p$). 

From \eqn{XAR}, \eqn{someXA^n} it is clear that
\bal
\Tr(X_AR)&=stR=\sum_{p\in{2\bZ+1}}\frac{\kappa}{\kappa+\ch p}\,,
\\
\Tr(X_AR)^2&=st\Tr(RR_++RR_-)+(st)^2\Tr(R^2)
\\&
=\sum_{p\in2\bZ+1}\frac{\kappa}{\kappa+\ch p}\left(\frac{2st\kappa}{\kappa+\ch(p+2)}
+\frac{(st)^2\kappa}{\kappa+\ch p}\right).
\eal
The last identity uses that the trace is invariant under shifts of all the subscripts 
of $R_n$ and under an overall change of sign, since the sum is over all odd $p$ 
and the function $R$ is even.

More generally, the three terms in $X_AR$ \eqn{XAR} can be thought of as three possible steps 
in a one-dimensional random walk where the step $e^{2i\alpha}$ from position $l$ is 
weighted by $sR_l$ and increases $l$ by one. Likewise $e^{-2i\alpha}$ decreases $l$ by 
one and is weighted by $tR_l$ and staying in place is weighted by 
$stR_l$. Finally, the trace enforces that the endpoint has $l=0$, so 
$\Tr(X_AR)^n$ is represented by this closed $n$-step random walk. The next few 
powers are
\bal
\Tr(X_AR)^3
&=6(st)^2\Tr(R^2R_+)+(st)^3\Tr(R^3)\,,
\\
\Tr(X_AR)^4
&=(st)^2\Tr(2R^2R_+^2+4RR_+^2R_{++})
+(st)^3\Tr(8R^3R_++4R^2R_+^2)
+(st)^4\Tr(R^4)\,,
\eal
Using this, one finally gets
\beq
\label{det-exp-A}
\det(1+X_AR)
=\exp\Tr\bigg[st\big(R-RR_+\big)
-\frac{(st)^2}{2}\left(R^2-4R^2R_++R^2R_+^2+2RR_+^2R_{++}\right)
+\cO\big((st)^3\big)\bigg]
\eeq
The terms up to order $(st)^5$ are given in Appendix~\ref{app:asym}.

\subsection{Symmetric representations}

The analog of equation \eqn{anti-gen} for the symmetric representation is
\bal
\label{sym-gen}
\prod_{j=1}^N\frac{1}{\left(1-se^{2i\alpha_j}\right)\left(1-te^{-2i\alpha_j}\right)}
&=1+\sum_{j=1}^N\left(se^{2i\alpha_j}+te^{-2i\alpha_j}\right)
+st\sum_{j,k=1}^Ne^{2i(\alpha_j-\alpha_k)}
\\&\quad{}
+\sum_{j\leq k}\left(s^2e^{2i(\alpha_j+\alpha_k)}+t^2e^{-2i(\alpha_j+\alpha_k)}\right)
+\cdots\,.
\eal
The subtle difference between the symmetric and antisymmetric representations 
is just the limit on the sum on the second line.

defining the density operator for the symmetric representations as
\beq
\rho_S\equiv\frac{1}{(1-se^{2i\alpha})(1-te^{-2i\alpha})}\rho_0
=\frac{1}{1-st}\left(\frac{1}{1-se^{2i\alpha}}+\frac{1}{1-te^{-2i\alpha}}-1\right)\rho_0\,.
\eeq
then the generating function of the Wilson loops in the symmetric representations in the 
grand canonical ensemble is
\beq
\label{grand-sym}
1+\sum_{k=1}^\infty(st)^kW_{\text{sym}^k}(\kappa)
=\frac{\det\left(1+\kappa\rho_S\right)}{\det\left(1+\kappa\rho_0\right)}
=\det\left(1+X_SR\right)\,,
\eeq
where
\beq
X_S=
\frac{1}{1-st}\left(\frac{se^{2i\alpha}}{1-se^{2i\alpha}}
+\frac{te^{-2i\alpha}}{1-te^{-2i\alpha}}+st\right)
=\frac{1}{1-st}\bigg(st+\sum_{j=1}^\infty
\left(s^je^{2ij\alpha}+t^je^{-2ij\alpha}\right)\bigg).
\eeq

The determinant can be written again as the exponent of the trace of the log 
of the argument, which requires to calculate traces of the form
\beq
\Tr\left(X_SR\right)^n
=\frac{1}{(1-st)^n}\Tr\bigg(stR+\sum_{j=1}^\infty
\left(s^je^{2ij\alpha}R+t^je^{-2ij\alpha}R\right)\bigg)^n.
\eeq
This is again a closed $n$-step random walk with arbitrary integer size 
steps weighted by an overall $1/(1-st)^n$ and then $s^jR_j$ for a $j$ 
size step to the right from position $l$, by $t^jR_l$ when going left and 
$stR_l$ when staying in the $l$\textsuperscript{th} position. 
For the first few $n$ this gives
\bal
n&=1:&&\frac{stR}{1-st}\,,
\\
n&=2:&&\frac{1}{(1-st)^2}\bigg((st)^2R^2+2\sum_{j=1}^\infty (st)^jRR_j\bigg),
\\
n&=3:&&\frac{1}{(1-st)^3}\bigg((st)^3R^3+6\sum_{j=1}^\infty (st)^{j+1}R^2R_j
+6\sum_{j=2}^\infty\sum_{l=1}^{j-1}(st)^jRR_lR_j\bigg),
\\
n&=4:&&\frac{1}{(1-st)^4}\bigg((st)^4R^3
+4\sum_{j=1}^\infty (st)^{j+2}\left(2R^3R_j+R^2R_j^2\right)
\\
&&&+8\sum_{j=2}^\infty\sum_{l=1}^{j-1}(st)^{j+1}
\left(2R^2R_lR_j+RR_l^2R_j\right)
+8\sum_{j=3}^\infty\sum_{l=1}^{j-2}\sum_{m=1}^{j-l-1}(st)^jRR_mR_lR_j
\\&&&
+2\sum_{j=2}^\infty\sum_{l,m=1}^{j-1}(st)^j \left(2RR_lR_mR_j+ RR_lR_{l-m}R_{j-m}\right)
\bigg).
\eal
Then\footnote{%
The terms of order $(st)^3$ and $(st)^4$ are in Appendix~\ref{app:sym}.}
\bal
\label{det-exp-S}
\det(1+X_SR)
&=\exp\left[\sum_{n=1}^\infty\frac{(-1)^{n+1}}{n}\Tr(X_SR)^n\right]
\\&=\exp\Tr\bigg[st\big(R-RR_+\big)
-\frac{(st)^2}{2}\big({-2}R+4RR_++R^2+2RR_{++}
\\&\hskip2cm{}
-4R^2R_+-4RR_+R_{++}
+R^2R_+^2+2RR_+^2R_{++}
\big)+\cO\big((st)^3\big)\bigg].
\eal
Clearly the term linear in $st$ is the same as in \eqn{det-exp-A}, giving again the 
fundamental representation. The next term, related to the first symmetric representation 
is very different from that in \eqn{det-exp-A}.

As seen above, for either symmetric or antisymmetric representations, 
the only nonzero terms in the expansion have equal powers of 
$s$ and $t$. It is therefore unambiguous to set $s=t$. If one further defines 
$s=t=e^{\sigma}$ then the densities are
\beq
\rho_A
=4e^{\sigma}\cosh^2\bigg(\frac{\sigma+2i\alpha}{2}\bigg)\rho_0\,,
\qquad
\rho_S
=\frac{e^{-\sigma}}{4\sinh^2\big(\frac{\sigma+2i\alpha}{2}\big)}\,\rho_0\,.
\eeq
Those insertions are rather reminiscent of the contributions due to fundamental 
matter fields in the Fermi-gas approach to 3d Chern-Simons-matter theories. 
Possibly some of the techniques employed there, like Wigner's phase space, 
could be used here as well, rather than the explicit commutators employed above. 
This may allow to study arbitrary representations more efficiently.

It should also be noted that in the case of ABJ(M) it was very useful to consider 
hook representations, whose generating functions is 
$(1+s_1 e^{2i\alpha})/(1-s_2e^{2i\alpha})$ \cite{Hatsuda:2013yua}. 
As seen above, the symmetric and antisymmetric representations are complicated 
enough. One reason is that the Index vanishes unless two 
loops of conjugate representations are introduced, at opposite poles of $S^3$.

\section{Large $N$ limit}
\label{sec:largeN}

To study the index in the large $N$ limit one can consider the grand canonical ensemble 
at large $\kappa$, as a saddle point equation guarantees that these limits are 
equivalent. After evaluating the large $\kappa=e^\mu$ expression, one gets the index by the 
integral transform ({\em c.f.}, \eqn{grand-R})
\beq
\label{inverse}
\cI_R(N)=\frac{1}{Z(N)}\int_{0}^{2i\pi} \frac{d\mu}{2\pi i}\, e^{-\mu N}\Xi(e^{\mu})W_R(e^\mu)\,.
\eeq

For the fundamental representation the term linear in $st$ in \eqn{det-exp-A} gives
\beq
\label{fund-kappa}
W_\Box(\kappa)
=\Tr(R-RR_+)
=\sum_{p\in2\bZ+1}\frac{\kappa\ch(p+2)}{(\kappa+\ch p)(\kappa+\ch(p+2))}\,.
\eeq
For large $\kappa$ one can use the continuum approximation for the sum
\bal
\label{integral}
W_\Box(\kappa)
&\sim
\half\int_{-\infty}^\infty dp\,\frac{\kappa(q^{-p/2-1}+q^{p/2+1})}
{(\kappa+q^{-p/2}+q^{p/2})(\kappa+q^{-p/2-1}+q^{p/2+1})}
\\&
=\frac{\kappa}{\big(\kappa^2-\frac{(1+q)^2}{q}\big) \log q}
\left[\frac{\kappa(1+q) \log \frac{\kappa+q^{-p/2}+q^{p/2}}
{q(\kappa+q^{-p/2-1}+q^{p/2+1})}}{2 (1-q)}
+\frac{\kappa^2\arctanh\frac{\sqrt{\kappa^2-4}}{\kappa+2q^{p/2+1}}}{\sqrt{\kappa^2-4}}
\right.\\&\quad
\left.{}
-\frac{\big(\kappa^2-2\frac{(1+q)^2}{q}\big) 
\arctanh\frac{\sqrt{\kappa^2-4}}{\kappa +2 q^{p/2}}}
{\sqrt{\kappa^2-4}}
\right]_{-\infty}^\infty
=\frac{1+q}{1-q}\,
\frac{1
+2\,\frac{(1-q^2)}{q\log q}\,
\frac{\arctanh\sqrt{1-4/\kappa^2}}{\kappa^2\sqrt{1-4/\kappa^2}}}
{1-\frac{(1+q)^2}{q\kappa^2}}\,.
\eal

At leading order at large $\kappa$ this is simply
\beq
W_\Box(\kappa)=\frac{1+q}{1-q}+\cO\big(\kappa^2\log\kappa\big).
\eeq
Since this leading asymptotics has no $\kappa$ dependence, 
the factor of $W_R(e^\mu)$ in \eqn{inverse} can be taken out of the integral, which 
just gives $Z(N)$ ({\it c.f.}, \eqn{grandpartition}). 
Hence at leading order at large $N$ (or $\kappa$) the Wilson loop 
in the canonical and grand-canonical ensembles are equal
\beq
\label{fund-largeN}
\cI_\Box(N)\sim W_\Box(\kappa)\sim\frac{1+q}{1-q}\,.
\eeq

One can consider higher dimensional representations. At order $(st)^2$ in 
\eqn{det-exp-A}, using the same integral approximation as for the fundamental, 
one finds
\beq
\label{two-box}
-\frac{1}{2}\Tr\big(R^2-4R^2R_++R^2R_+^2+2RR_+^2R_{++}\big)
\sim\frac{1}{2}\frac{1+q^2}{1-q^2}\,.
\eeq
Expanding the exponent in \eqn{det-exp-A} gives
\beq
W_{\scalebox{.6}{\tiny\yng(1,1)}}(\kappa)
\sim\frac{1}{2}\left(\frac{(1+q)^2}{(1-q)^2}+\frac{1+q^2}{1-q^2}\right),
\eeq
and at this order this is also the asymptotic value of 
$\cI_{\scalebox{.6}{\tiny\yng(1,1)}}(N)$.

Repeating the calculation for the term cubic in $st$ in \eqn{det-exp-A} gives at leading order
\beq
\frac{1}{3}\frac{1+q^3}{1-q^3}+O\big(\kappa^{-2}\big)\,.
\eeq
so to leading order the Wilson loop in either the canonical or grand canonical 
ensemble is simply
\beq
W_{\scalebox{.5}{\tiny\yng(1,1,1)}}(\kappa)
\sim
\cI_{\scalebox{.5}{\tiny\yng(1,1,1)}}(N)
\sim
\frac{1}{6}\left(\frac{(1+q)^3}{(1-q)^3}+3\frac{1+q^2}{(1-q)^2}+2\frac{1+q^3}{1-q^3}\right).
\eeq

Indeed, the leading large $\kappa$ behavior for all terms in \eqn{det-exp-A} 
up to $(st)^5$ (see Appendix~\ref{app:asym}) is
\beq
\frac{(st)^n}{n}\frac{1+q^n}{1-q^n}\,.
\eeq
It is natural to conjecture that this pattern continues, so that at large $\kappa$ the determinant in 
\eqn{det-exp-A} would be
\bal
\det(1+X_AR)
&\sim\exp\left[\sum_{n=1}^\infty\frac{(st)^n}{n}\frac{1+q^n}{1-q^n}\right]
=\exp\sum_{n=1}^\infty\frac{(st)^n}{n}\left(\frac{2}{1-q^n}-1\right)
\\&
=(1-st)\exp\left[2\sum_{n=1}^\infty\sum_{k=0}^\infty\frac{(stq^k)^n}{n}\right]
=\frac{1-st}{\prod_{k=0}^\infty(1-stq^k)^2}\,.
\eal
These products are known as $q$-Pochhammer symbols
\beq
\det(1+X_AR)=\frac{1+st}{(st;q)_\infty^2}
=\frac{1}{(st;q)_\infty(stq;q)_\infty}\,.
\eeq
It is simple to expand these expressions to arbitrary orders in $st$.

Applying the same techniques to the symmetric representation, the leading order at large 
$\kappa$ for the term multiplying $(st)^2$ in \eqn{det-exp-S} is
\beq
\frac{1}{2}\frac{1+q^2}{1-q^2}\,,
\eeq
which is the same as the antisymmetric representation \eqn{two-box}. Indeed the 
same is true for the three and four box symmetric representations (see Appendix~\ref{app:sym}). 
Again it would be natural to conjecture that this pattern continues for all $N$ and that 
also for the symmetric representation
\beq
\det(1+X_SR)
\sim\frac{1-st}{\prod_{k=0}^\infty(1-stq^k)^2}
=\frac{1}{(st;q)_\infty(stq;q)_\infty}\,.
\eeq

Having seen that the answer at large $N$ for all the the above examples 
does not depend on whether they were symmetric or antisymmetric representation, 
it is natural to further conjecture that the leading result depends only on the 
number of boxes in the Young diagram.%
\footnote{Alternatively, on the length of the longest hook in the diagram, which for the 
antisymmetric and symmetric representations is equal to the total number of boxes.}
 
\begin{figure}[ht]
\centering
\epsfig{file=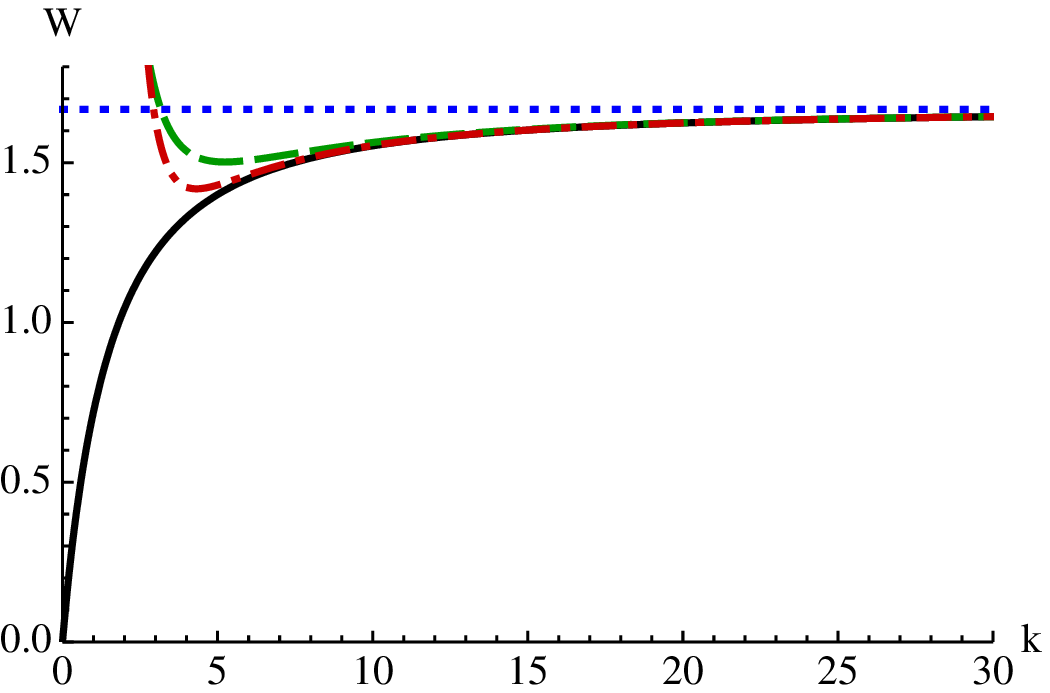,width=7.5cm
\psfrag{k}{\small$\kappa$}
\psfrag{W}{\hskip-3mm\small$W_\Box$}
\psfrag{0.0}{\raisebox{-.3mm}{\footnotesize$\;0$}}
\psfrag{0.5}{\raisebox{-.3mm}{\footnotesize$\!\!\!0.5$}}
\psfrag{1.0}{\raisebox{-.3mm}{\footnotesize$\!\!\!1.0$}}
\psfrag{1.5}{\raisebox{-.3mm}{\footnotesize$\!\!\!1.5$}}
\psfrag{0}{\raisebox{-1mm}{\footnotesize$0$}}
\psfrag{5}{\raisebox{-1mm}{\footnotesize$5$}}
\psfrag{10}{\raisebox{-1mm}{\footnotesize$10$}}
\psfrag{15}{\raisebox{-1mm}{\footnotesize$15$}}
\psfrag{20}{\raisebox{-1mm}{\footnotesize$20$}}
\psfrag{25}{\raisebox{-1mm}{\footnotesize$25$}}
\psfrag{30}{\raisebox{-1mm}{\footnotesize$30$}}
}\qquad
\epsfig{file=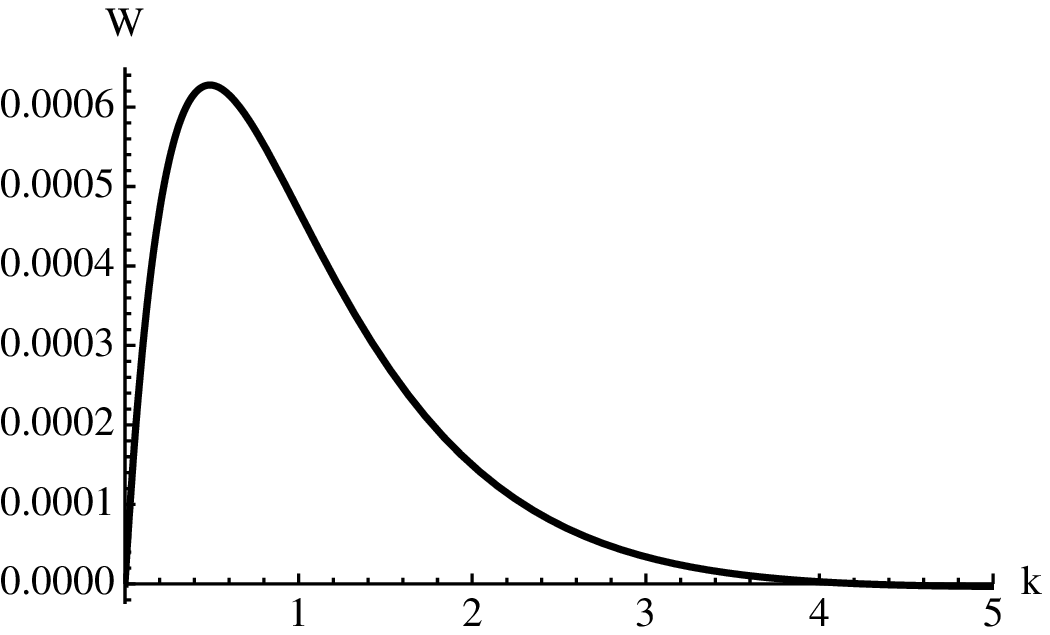,width=8cm
\psfrag{k}{\small$\kappa$}
\psfrag{W}{\hskip-5mm\small Integral approximation of $W_\Box$ minus exact sum}
\psfrag{0.0000}{\raisebox{-.3mm}{\footnotesize$\quad\ \,0$}}
\psfrag{0.0001}{}
\psfrag{0.0002}{\raisebox{-.3mm}{\footnotesize$\!\!\!\!\!0.0002$}}
\psfrag{0.0003}{}
\psfrag{0.0004}{\raisebox{-.3mm}{\footnotesize$\!\!\!\!\!0.0004$}}
\psfrag{0.0005}{}
\psfrag{0.0006}{\raisebox{-.3mm}{\footnotesize$\!\!\!\!\!0.0006$}}
\psfrag{1}{\raisebox{-1mm}{\footnotesize$1$}}
\psfrag{2}{\raisebox{-1mm}{\footnotesize$2$}}
\psfrag{3}{\raisebox{-1mm}{\footnotesize$3$}}
\psfrag{4}{\raisebox{-1mm}{\footnotesize$4$}}
\psfrag{5}{\raisebox{-1mm}{\footnotesize$5$}}
}
\caption{\small The continuum approximation of $W_\Box(\kappa)$ for $q=1/4$: 
On the left graph different approximations of $W_\Box(\kappa)$ are plotted as function 
of $\kappa$. The dotted blue line is the asymptotic value $\frac{1+q}{1-q}=5/3$. 
The green dashed line includes the $1/\kappa^2$ correction and the red dash-dotted line 
also the $1/\kappa^4$ corrections in \eqn{large-kappa}. 
The full continuum approximation in \eqn{integral} and 
the numerical evaluation of the sum in \eqn{fund-kappa} are the solid black line. 
The difference between them cannot be seen at this 
scale, and is plotted in the right graph.
\label{fig:approx}}
\end{figure}

The discussion so far involved only the leading order at large $N$ of $\cI_R(N)$. 
To find subleading corrections it is possible to examine the corrections to 
$W_R(\kappa)$. For the fundamental representation one can expand equation 
\eqn{integral} to find
\bal
\label{large-kappa}
W_\Box(\kappa)
&\sim
\frac{1+q}{1-q}
+\frac{(1+q)^2}{q \kappa ^2} \left(\frac{1+q}{1-q}+2 \log_q\kappa\right)
\\&\quad{}+
\frac{(1 + q)^2}{q^2\kappa^4}
\left(\frac{(1 + q)^3}{1-q}-\frac{2q}{\log q}+2(1+4q + q^2)\log_q\kappa\right)
+\cO(\kappa^{-6}\log\kappa)\,.
\eal
As can be seen in Fig.~\ref{fig:approx}, for $q=1/4$ the continuum approximation 
\eqn{integral} of the sum \eqn{fund-kappa} is very good, 
differing from the exact expression by less than 1\% for arbitrary $\kappa$ and much 
much better for large $\kappa$.  A wide range of $q$ gives similarly looking graphs. 
The same figure also shows the large $\kappa$ expansion of this result, which 
also furnishes a good approximation.

The continuum approximation breaks down, though, for negative $\kappa$, where 
\eqn{integral} and \eqn{large-kappa} have branch cuts. The origin of this is the infinite 
number of poles of $W_\Box(\kappa)$ at $\kappa=-q^{p/2}-q^{-p/2}$ \eqn{fund-kappa}. 
These poles all occur at zeros of $\Xi(\kappa)$ \eqn{SUNXi}, so they have no dramatic 
effect on the integral \eqn{inverse}. The branch cuts, on the other hand, make the 
integral ambiguous. It would be interesting to find a workaround to extract the 
corrections to $\cI_\Box(N)$ beyond the leading large $N$ result 
\eqn{fund-largeN}.

\section{Fundamental representation at small $N$}
\label{sec:finiteN}

To evaluate the index with Wilson loops at finite $N$, one can also 
expand \eqn{fund-kappa} in a power series in $\kappa$
\beq
W_\Box(\kappa)
=\sum_{N=1}^\infty(-1)^{N+1}\kappa^N\sum_{p\in2\bZ+1}\sum_{k=1}^N\frac{1}{\ch^kp\ch^{N-k}(p+2)}\,.
\eeq
Multiplying by $\Xi(\kappa)$ \eqn{grandpartition} one finds at order $\kappa^N$
\bal
Z(N)\cI_\Box(N)=\sum_{n=1}^N
Z(N-n)(-1)^{n+1}\sum_{p\in2\bZ+1}\sum_{k=1}^n\frac{1}{\ch^kp\ch^{n-k}(p+2)}\,.
\eal

Using \eqn{SUNXi} the resulting sums are
\beq
Z(1)=\sum_{p\in2\bZ+1}\frac{1}{\ch p}\,,
\qquad
Z(2)=
\frac{1}{2}\left(\bigg(\sum_{p\in2\bZ+1}\frac{1}{\ch p}\bigg)^2
-\sum_{p\in2\bZ+1}\frac{1}{\ch^2p}\right).
\eeq
These sums can be calculated (see the techniques in \cite{Zucker1979, BDF1, BDF2})
\beq
Z(1)=\frac{kK}{\pi}\,,
\quad
\sum_{p\in2\bZ+1}\frac{1}{\ch^2p}=\frac{KE-(1-k^2)K^2}{\pi^2}\,,
\quad
\sum_{p\in2\bZ+1}\frac{1}{\ch p\ch(p+2)}=\frac{q}{1-q^2}\,,
\eeq
where $K$ and $E$ are complete elliptic integrals with modulus 
$k=\theta_2^2/\theta_3^2$.

This gives
\beq
\cI_\Box(2)
=\frac{1}{Z(2)}\left(\frac{k^2K^2}{\pi^2}
-\frac{KE-(1-k^2)K^2}{\pi^2}
-\frac{q}{1-q^2}\right)
=2
-\frac{1}{Z(2)}\frac{q}{1-q^2}\,,
\eeq
where \cite{BDF1}
\beq
Z(2)=\frac{K(K-E)}{2\pi^2}\,.
\eeq
One can consider higher dimensional representations and $N>2$, but the necessary 
sums get rather unwieldily.

\section*{Acknowledgements}
It is a pleasure to thank Jun Bourdier and Jan Felix for related collaboration and interesting 
discussions.
This research is underwritten by an STFC advanced fellowship.

\appendix

\section{Theta functions}
\label{app:theta}


The Jacobi theta function $\vartheta_3(z,q)$ is given by the series 
and product representations as
\beq
\label{theta3}
\vartheta_3(z,q)=\sum_{n=-\infty}^\infty q^{n^2}e^{2inz}
=\prod_{k=1}^\infty\big(1-q^{2k}\big)
\big(1+2q^{2k-1}\cos(2z)+q^{4k-2}\big),
\eeq
in terms of which the auxiliary theta functions are given by 
\bal
\label{auxtheta}
\vartheta_1(z,q)&=iq^{1/4}e^{-iz}\vartheta_3\big(z-\tfrac{\pi\tau}{2}-\tfrac{\pi}{2},q\big) 
\\
\vartheta_2(z,q)&=q^{1/4}e^{-iz}\vartheta_3\big(z-\tfrac{\pi\tau}{2},q\big)
\\
\vartheta_4(z,q)&=\vartheta_3(z-\tfrac{\pi}{2},q)\,.
\eal

\section{Higher order expansions}
\label{app:higher}
\subsection{Antisymmetric representations}
\label{app:asym}

The next few terms in \eqn{det-exp-A} which could fit on less than a 
page in a reasonable font are
{\small
\bal
&
\det(1+X_AR)=\exp\Tr\bigg[
st\big(R-RR_1\big)
-\frac{(st)^2}{2} \big(R^2R_1^2+R^2-4RR_1^2+2RR_1^2 R_2\big)
\\&\quad{}
+\frac{(st)^3}{3}
\big({-}R^3R_1^3
+R^3
+6R^2R_1^3
-3R^2R_1^2
-6RR_1^3
-6RR_1^3 R_2^2
+6RR_1^2 R_2^2
+6RR_1^3 R_2
-3RR_1^2 R_2^2 R_3\big)
\\&\quad{}
-\frac{(st)^4}{4}
\big(R^4R_1^4
+R^4-8R^3R_1^4
+8R^3R_1^3
+12R^2R_1^4
-8R^2R_1^3
+6R^2R_1^4R_2^2
-8R^2R_1^3 R_2^2
+4R^2R_1^2 R_2^2
\\&\qquad{}
-8RR_1^4
+8RR_1^4 R_2^3
-16RR_1^3 R_2^3
+8RR_1^2 R_2^3
-24RR_1^4 R_2^2
+16RR_1^3 R_2^2
+8RR_1^2 R_2^3R_3^2
-8RR_1^2 R_2^2 R_3^2
\\&\qquad{}
+12RR_1^4 R_2
+8RR_1^3 R_2^3 R_3
-16 R_1^2R_2^3 R_3
+4RR_1^2 R_2^2 R_3^2 R_4\big)
\\&\quad{}
+\frac{(st)^5}{5} \big({-}R^5R_1^5
+R^5
+10R^4R_1^5
-15R^4R_1^4
-20R^3R_1^5
+30R^3R_1^4
-5R^3R_1^3
+20R^2R_1^5
-10R^2R_1^4
\\&\qquad{}
-20R^2R_1^5 R_2^3
+40R^2R_1^4 R_2^3
-30R^2R_1^3 R_2^3
+10R^2R_1^2 R_2^3
+30R^2R_1^5 R_2^2
-30R^2R_1^4 R_2^2
+10R^2R_1^3 R_2^2
\\&\qquad{}
-5R^2R_1^3 R_2^3 R_3^2
+10R^2R_1^2 R_2^3 R_3^2
-5R^2R_1^2 R_2^2 R_3^2
-10RR_1^5
-10RR_1^5 R_2^4
+30RR_1^4 R_2^4
-30RR_1^3R_2^4
\\&\qquad{}
+10RR_1^2 R_2^4
+40RR_1^5 R_2^3
-60 RR_1^4 R_2^3
+20 RR_1^3 R_2^3
-10 RR_1^2 R_2^4 R_3^3
+20 RR_1^2 R_2^3 R_3^3
-10 RR_1^2 R_2^2 R_3^3
\\&\qquad{}
-60RR_1^5 R_2^2
+30R R_1^4 R_2^2
-30 RR_1^3 R_2^4 R_3^2
+30 RR_1^2 R_2^4 R_3^2
+40 RR_1^3 R_2^3 R_3^2
-20 RR_1^2 R_2^3 R_3^2
\\&\qquad{}
-20 RR_1^3 R_2^2 R_3^2
-10RR_1^2 R_2^2 R_3^3 R_4^2
+10R R_1^2 R_2^2 R_3^2 R_4^2
+20 RR_1^5 R_2
-15RR_1^4 R_2^4 R_3
+60 RR_1^3 R_2^4 R_3
\\&\qquad{}
-30 RR_1^2 R_2^4 R_3
-20 RR_1^3 R_2^3R_3
-20 RR_1^2 R_2^3 R_3^3 R_4
+20 RR_1^2 R_2^2 R_3^3 R_4
+10 RR_1^2 R_2^3R_3^2 R_4
\\&\qquad{}
-5 RR_1^2 R_2^2 R_3^2 R_4^2 R_5\big)
+\cO\big((st)^6\big)\bigg].
\eal
}

\subsection{Symmetric representations}
\label{app:sym}

The next few terms in \eqn{det-exp-S} are
{\small
\bal
&\hskip-4mm
\det(1+X_SR)
=\exp\Tr\bigg[
st\big(R-R R_1\big)
\\&{}
-\frac{(st)^2}{2} \big(
R^2R_1^2
-4R^2R_1
+R^2
+4RR_1
+2 RR_1^2 R_2
-4 RR_1 R_2
+2RR_2-2 R\big)
\\&{}
+\frac{(st)^3}{3} \big(
{-}R^3R_1^3
+6R^3R_1^2
-6R^3R_1
+R^3
-9R^2R_1^2
+18R^2R_1
-6R^2R_1^3 R_2
+18R^2R_1^2 R_2
\\&\quad{}
-18R^2R_1 R_2
+6R^2R_2
-3 R^2
-9RR_1
+6RR_1^3 R_2
-18RR_1^2 R_2
+18RR_1 R_2
-6R R_2
\\&\quad{}
-6R R_1^2 R_3
-3R R_1^2 R_2^2 R_3
+12R R_1 R_3
+12R R_1^2 R_2 R_3
-12R R_1R_2 R_3
-3R R_3
+3 R\big)
\\&{}
-\frac{(st)^4}{4} \big(
R^4R_1^4
-8R^4 R_1^3
+12R^4R_1^2
-8R^4R_1
+R^4
+16R^3R_1^3
-48R^3R_1^2
+32R^3R_1
+8R^3R_1^4 R_2
\\&\quad{}
-32R^3R_1^3 R_2
+48R^3R_1^2 R_2
-32R^3R_1 R_2
+8R^3R_2
-4 R^3
+36R^2 R_1^2
+6 R^2R_1^4R_2^2
-24 R^2R_1^3 R_2^2
\\&\quad{}
+36R^2 R_1^2 R_2^2
-24R^2 R_1 R_2^2
+6R^2 R_2^2
-48R^2 R_1
-24R^2 R_1^4 R_2
+96R^2 R_1^3 R_2
-144R^2 R_1^2 R_2
\\&\quad{}
+96R^2 R_1 R_2
-24 R^2R_2
+8R^2 R_1^3 R_3
-24R^2 R_1^2 R_3
+8R^2 R_1^3 R_2^2 R_3
-24R^2R_1^2 R_2^2 R_3
\\&\quad{}
+24R^2 R_1 R_2^2 R_3
-4R^2 R_2^2 R_3
+24R^2 R_1 R_3
-16R^2R_1^3 R_2 R_3
+48R^2 R_1^2 R_2 R_3
-48R^2 R_1 R_2 R_3
\\&\quad{}
+16R^2 R_2 R_3
-8R^2R_3
+6 R^2
-4R R_1^2 R_3^2
+16R R_1
+12R R_1^4 R_2
-48 RR_1^3 R_2
+72 RR_1^2R_2
\\&\quad{}
-48R R_1 R_2
+12 RR_2
-16 RR_1^3 R_3
+8R R_1^3 R_2^3 R_3
+48 RR_1^2 R_3
-48 RR_1^3 R_2^2 R_3
+72 RR_1^2 R_2^2 R_3
\\&\quad{}
-48R R_1 R_3
+48 RR_1^3 R_2 R_3
-144 RR_1^2 R_2 R_3
+72R R_1 R_2 R_3
+8R R_3
+8 RR_1^2 R_4
+8R R_1^2 R_2^2R_4
\\&\quad{}
-16 RR_1 R_2^2 R_4
+4R R_2^2 R_4
+4 RR_1^2 R_3^2 R_4
+4R R_1^2 R_2^2R_3^2 R_4
-8 RR_1^2 R_2 R_3^2 R_4
-16 RR_1 R_4
\\&\quad{}
-16R R_1^2 R_2 R_4
+32 RR_1R_2 R_4
-8R R_2 R_4
-16 RR_1^2 R_3 R_4
-16 RR_1^2 R_2^2 R_3 R_4
+16R R_1R_2^2 R_3 R_4
\\&\quad{}
+16 RR_1 R_3 R_4
+32R R_1^2 R_2 R_3 R_4
-32 RR_1 R_2 R_3 R_4
+4R R_4
-4 R\big)
+\cO\big((st)^5\big)\bigg].
\eal
}
\bibliography{references}

\end{document}